\documentclass{appolb}
\usepackage{graphicx}
\usepackage[numbers,sort&compress]{natbib}

\newcommand{\cc}{\mbox{COSY--11}}
\begin{document}
\title{STUDY OF THE NN$\eta'$ PRODUCTION WITH \cc%
\thanks{Presented at the II International Symposium on Mesic Nuclei, Krak\'ow, Poland,
September 22–25, 2013.}%
}
\author{Eryk Czerwi\'nski$^1$, Pawe\l{} Moskal$^{1,2}$ and Micha\l{} Silarski$^1$\\for the \cc\ Collaboration
\address{$^1$ Institute of Physics, Jagiellonian University,\\ul. Reymonta 4, 30-059 Cracow, Poland}
\address{$^2$ Institute of Nuclear Physics, Research Centre J\"{u}lich,\\ Leo-Brandt-Stra\ss e, 52428 J\"{u}lich, Germany}
}
\maketitle
\begin{abstract}
We describe a new high precision measurement of the production cross-section for 
the $\eta'$ meson in
proton-proton collisions $\sigma_{pp\to pp\eta'}$ for five beam momenta at low access energy region $Q$
conducted at the \cc\ detection system
together with an updated results of all other previous measurements of $\sigma_{pp\to pp\eta'}$ at \cc.
\end{abstract}
\section{Introduction}
Recently the increased interest in the properties of the $\eta$ and $\eta'$ meson
can be observed due to extensive experimental search of the $\eta$ and $\eta'$ bound states performed e.g. at
  COSY~\cite{COSY11-MoskalSmyrski,WASA-at-COSY-SkuMosKrze,Adlarson2013,GEM,3Heeta-PL-Smyrski,3Heeta-Mersmann},
  ELSA~\cite{cbelsa},
  GSI~\cite{eta-prime-mesic-GSI-tanaka,eta-prime-mesic-GSI-Itahashi},
  JINR~\cite{JINR}, 
  JPARC~\cite{eta-mesic-JPARC-Fujioka,eta-mesic-JPARC-Fujioka-Itahashi},
  LPI~\cite{LPI}, and
  MAMI~\cite{ELSA-MAMI-plan-Krusche,gamma3He-Pheron} as well as intensive 
theoretical
investigations e.g.~\cite{wilkin2,bass,eta-prime-mesic-Nagahiro,Mesic-Bass,EtaMesic-Hirenzaki,eta-prime-mesic-Hirenzaki,eta-prime-mesic-Nagahiro-Oset,
ETA-Friedman-Gal,ETA-Gal-Cieply,Wycech-Acta,WYCECHGREE-GW-zGAla,Mesic-Kelkar,Bass10Acta}.

Properties of $\eta'$ in nuclear medium are related with the effects of $U_A(1)$ anomaly at finite 
density~\cite{EtaMesic-Hirenzaki,bass,Mesic-Bass,eta-prime-mesic-Nahagiro},
which is reflected in the large mass of the $\eta'$ meson compared to the masses of the other members of the 
pseudoscalar meson nonet~\cite{klimt,jido}, and with the $\eta$-$\eta'$ mixing~\cite{Mesic-Bass,Hirenzaki-mixing}.

\cc\ experiment~\cite{Brauksiepe,C11Klaja} has provided already an important data for these studies~\cite{Moskal1,Moskal2,Khoukaz,pklaja},
with the most precise direct measurement of the total width of the $\eta'$ meson $\Gamma_{\eta'}~$\cite{ErykPhD,eryk_prl},
and the first rough estimation of the $\eta'$-N interaction from the excitation function of the cross section for the $pp\to pp\eta'$ reaction~\cite{Swave}.
Here we describe an analysis of the data used earlier for $\Gamma_{\eta'}$ determination
in view of the extraction of the production cross-section for
the $\eta'$ meson $\sigma_{pp\to pp\eta'}$ in proton-proton collisions and an update of the
$\sigma_{pp\to pp\eta'}$ values presented previously~\cite{Moskal1,Moskal2,Khoukaz}.
\section{Experiment}
\begin{figure}[htb]
\vspace{-0.5cm}
\centerline{%
\includegraphics[width=0.6\textwidth]{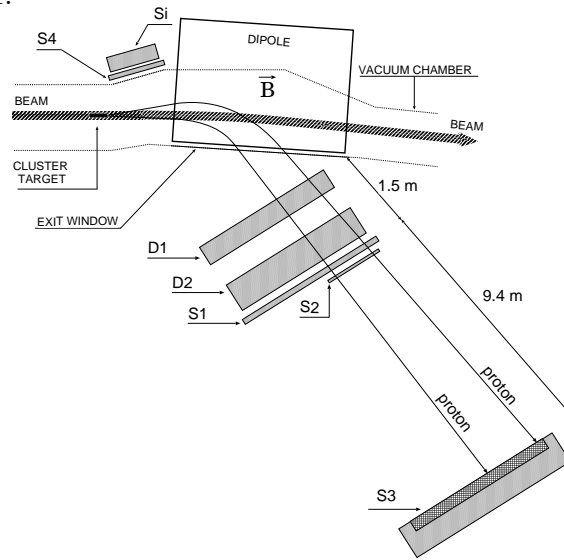}}
\caption{Schematic view of the COSY--11 detector setup (top view).
         S1, S2, S3 and S4 denote scintillator detectors, D1 and D2 indicate drift chambers and  Si stands
         for the silicon-pad detector.}
\label{fig:c11setup}
\end{figure}

In the reported measurement the $\eta'$ meson was produced in proton-proton collisions reaction and its
mass was reconstructed based on the momentum vectors of protons taking part in the
$pp~\to~pp\eta^{\prime}$ reaction which was measured at five different beam momenta
using the \cc\ detector setup~\cite{Brauksiepe,C11Klaja} installed at the cooler synchrotron COSY~\cite{Maier}
in Research Centre J\"ulich.
The schematic view of the \cc\ detector setup is presented in Fig.~\ref{fig:c11setup}.

The collision of a proton from the 
beam with a proton cluster target may cause an $\eta'$ meson creation.
In that case all outgoing nucleons have been registered by the \cc\ detectors, whereas
for the $\eta^{\prime}$ meson identification the missing mass technique was applied.
The COSY beam momentum and the dedicated zero degree \cc\ facility
enabled the measurement at an excess energy  down the fraction of an MeV above the kinematic threshold
for the $\eta^{\prime}$ meson production.
Modification of the \cc\ target system allowed to decrease effective beam momentum spread and
therefore enabled precise determination of the access energy $Q$ with the precision of 0.10 MeV.
Good control of the systematic uncertainties was possible due to measurement performed at five different values of $Q$
and monitoring of the beam and target properties~\cite{monitoring}.
On the other hand the achieved missing mass resolution in the order of the total width of the
$\eta'$ meson itself~\cite{eryk_prl} improved significantly the $\eta'$ production cross section measurement.
The number of registered $\eta'$ mesons was obtained from the missing mass spectra for each $Q$ value
and corrected for the detector geometrical acceptance and registration efficiency.
The luminosity value was determined using comparison of the cross section of
$pp\to pp$ reaction determined by EDDA Collaboration~\cite{edda2004} and the number of registered elastically scattered protons.
\section{Results}
Since $\sigma_{pp\to pp\eta'}$ measured at \cc\ was obtained with the
luminosity determination based on the EDDA data available that time~\cite{edda1997}, we updated these numbers accordingly to superseded data~\cite{edda2004}.
\cc\ measurement at $Q=16.4$~MeV~\cite{pklaja} was already reported with new EDDA data~\cite{edda2004}, whereas SPESIII~\cite{Hibou:1998de} and DISTO~\cite{Balestra:2000ic}
used different techniques for luminosity determination.
\begin{figure}[htb]
\centerline{%
\includegraphics[width=0.72\textwidth]{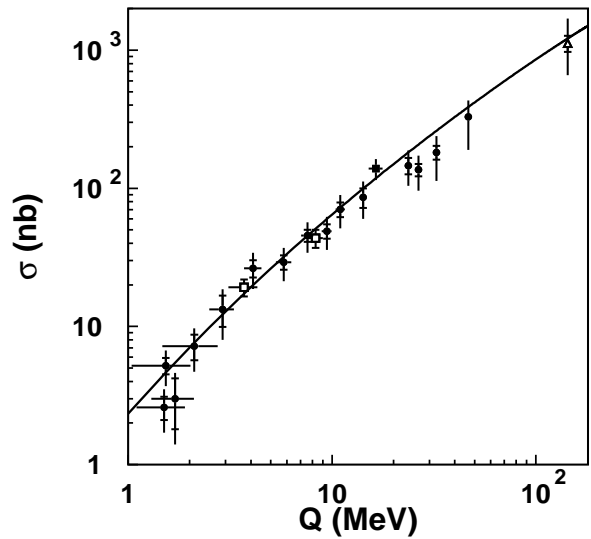}}
\vspace{-0.5cm}
\caption{Total cross section for the $\eta'$ meson production in the proton-proton collision as a function of the access energy $Q$
for the $pp\to pp\eta'$ reaction measured at: \cc\ (solid circles - updated values of~\cite{Moskal1,Moskal2,Khoukaz},
solid square - measurement~\cite{pklaja} with usage of EDDA 2004 data~\cite{edda2004}),
SPESIII (open squares)~\cite{Hibou:1998de}
and DISTO (open triangle)~\cite{Balestra:2000ic}. The solid line shows parametrization of the experimental data using formula~\ref{eq:colin}.}
\label{fig:cross}
\end{figure}
The experimental data presented at Fig.~\ref{fig:cross} are
compared to  the analytical parametrization derived  by F\"{a}ldt and
Wilkin~\cite{Faeldt:1996na,Faldt:1997jm}  which   takes  into  account
final state interaction of the protons:
\begin{equation}
\sigma_{pp\to pp\eta'}\left(Q\right)=C \frac{Q^2}{m_p p_{LAB}}\frac{1}{\left(1 + \sqrt{ 1 + \frac{Q}{\epsilon}}\right)^2},
\label{eq:colin}
\end{equation}
where $Q$ denotes the excess energy, $p_{LAB}$ beam momentum,  $m_p$
proton mass. The parameters $\epsilon=0.75^{+0.20}_{-0.15}$~MeV and $C=45^{+10}_{-9}$~mb denote the Coulomb distortion and
constant factor, respectively, and have been determined by fitting this formula to the experimental data.
Values of $pp\to pp\eta'$ cross sections determined at \cc\ are 
gathered in the Table~\ref{tab:updated} apart of the new measurement reported here, which is still in the final stage of the analysis.
\begin{table}[htb]
\begin{center}
\begin{tabular}{ r@{.}l c r@{.}l | r l l}
  \multicolumn{5}{c}{$Q~\left[MeV\right]$} &
  \multicolumn{3}{c}{$\sigma_{pp\to pp\eta'}~\left[nb\right]$} \\
  \noalign{\smallskip}
  \hline
  \hline
  \noalign{\smallskip}
 1&5 &$\pm$&0&4 &  2.6 &$\pm$~0.5 &$\pm$~0.4 \\
 1&53&$\pm$&0&49&  5.2 &$\pm$~0.7 &$\pm$~0.8 \\ 
 1&7 &$\pm$&0&4 &  3.0 &$\pm$~1.2 &$\pm$~0.5 \\ 
 2&11&$\pm$&0&64&  7.2 &$\pm$~1.5 &$\pm$~1.1 \\ 
 2&9 &$\pm$&0&4 & 13.3 &$\pm$~3.4 &$\pm$~2.0 \\ 
 4&1 &$\pm$&0&4 & 26.4 &$\pm$~3.8 &$\pm$~4.0 \\ 
 5&80&$\pm$&0&50& 29.2 &$\pm$~3.5 &$\pm$~4.4 \\ 
 7&57&$\pm$&0&51& 45.5 &$\pm$~4.5 &$\pm$~6.8 \\ 
 9&42&$\pm$&0&53& 49.0 &$\pm$~5.9 &$\pm$~7.4 \\ 
10&98&$\pm$&0&56& 70.5 &$\pm$~8.6 &$\pm$~11\\
14&21&$\pm$&0&57& 86   &$\pm$~14  &$\pm$~13\\
16&4 &$\pm$&1&3 &139   &$\pm$~3   &$\pm$~21\\
23&64&$\pm$&0&64&146   &$\pm$~20  &$\pm$~22\\
26&5 &$\pm$&1&0 &136   &$\pm$~14  &$^{+22}_{-26}$ \\
32&5 &$\pm$&1&0 &182   &$\pm$~21  &$^{+36}_{-48}$ \\
46&6 &$\pm$&1&0 &329   &$\pm$~18  &$^{+85}_{-122}$ 
\end{tabular}
\end{center}
  \caption{Updated values of production cross-sections for the $\eta'$ meson
	in proton-proton collisions measured at \cc\ detector~\cite{Moskal1,Moskal2,Khoukaz,pklaja} with statistical and systematic uncertainties, respectively.}
  \label{tab:updated}
\end{table}

\section{Acknowledgments}
This work has been supported by the Polish National Science Center through grants No.
0320/B/H03/2011/40, 2011/01/B/ST2/00431, 2011/03/\-B/ST2/01847,   
2011/01/D/ST2/00748, 2011/03/N/ST2/02652, by the Foundation for Polish Science through the
project HOMING PLUS BIS/2011-4/3,
by the European Commission under the 7th Framework Programme through the ‘Research
Infrastructures’ action of the ‘Capacities’ Programme (FP7-INFRA\-STRUCTURES-2008-1, Grant Agreement No. 227431) and
by the FFE grants of the Research Center J\"{u}elich.

\end{document}